\documentclass[twocolumn,amsmath,amssymb]{revtex4}
\usepackage{graphicx}
\usepackage[tight]{subfigure}
\usepackage{dcolumn}
\usepackage{bm}

\begin{document}
 
\title{The knee in the cosmic ray energy spectrum from the simultaneous EAS charged particles and muon density spectra}

\author {Biplab Bijay$^{1,a}$, Prabir Banik$^{1,b}$ and Arunava Bhadra$^{1,c}$}

\affiliation{$^{1}$ High Energy $\&$ Cosmic Ray Research Centre, University of North Bengal, Siliguri, West Bengal, India 734013\\
$^{a}$ biplabbijay@rediffmail.com, \\
$^{b}$ pbanik74@yahoo.com \\
$^{b}$ aru\_bhadra@yahoo.com }
       
\begin{abstract}
In this work we examine with the help of Monte Carlo simulation whether a consistent primary energy spectrum of cosmic rays emerges from both the experimentally observed total charged particles and muon size spectra of cosmic ray extensive air showers considering primary composition may or may not change beyond the knee of the energy spectrum. It is found that EAS-TOP observations consistently infer a knee in the primary energy spectrum provided the primary is pure unchanging iron whereas no consistent primary spectrum emerges from simultaneous use of the KASCADE observed total charged particle and muon spectra. However, it is also found that when primary composition changes across the knee the estimation of spectral index of total charged particle spectrum is quite tricky, depends on the choice of selection of points near the knee in the size spectrum.  
\end{abstract}

\maketitle

\section{Introduction}

The primary energy spectrum of all particle cosmic rays is known to exhibit a power law behavior with few features including a slight bend of the spectrum at about 3 PeV, the so called knee of the spectrum, where the power law spectral index changes from about -2.7 to nearly -3.0. The knee is generally believed to be of astrophysical origin. The common explanations of the knee include rigidity-dependent upper limit on the energy that cosmic ray protons can attain at supernova remnants \cite{t1}, leakage of cosmic rays from the galaxy \cite{t2}, a nearby single source \cite{t3}, mass distribution of progenitors of cosmic ray sources \cite{t4}  etc. 

The primary cosmic ray particles after entering into the Earth's atmosphere interact with the atmospheric nuclei and produce secondary particles. The detection of cosmic rays above the atmosphere is thus the only way to obtain direct measurements of the characteristics of primary cosmic ray particles including their energy spectra and mass composition. The energy spectrum of primary cosmic rays has been measured directly through satellite or balloon borne detectors up to few hundreds TeV. Above such energy direct methods for studying primary cosmic rays become inefficient due to sharp decrease in the flux of primary particles and the study of primary cosmic rays has to perform indirectly, through the observation of cosmic ray extensive air shower (EAS) which are cascades of secondary particles produced by interactions of cosmic ray particles with atmospheric nuclei. From their experimental results the Moscow State University group first noticed that the EAS electron size (total electron content) spectrum had a pronounced increase of slope ($\beta$ increases suddenly) at a size corresponding to a primary energy of about 3 PeV \cite{t5} which was inferred as due to a break or the knee in the cosmic ray primary energy spectrum. Since then many EAS experiments covering this energy range confirm such a break in the spectral index of electron size spectrum and the existence of the knee in the cosmic ray energy spectrum is now considered as a well- established fact. 

Some authors, however, cast doubt on the astrophysical origin of the knee. In particular a new type of interaction that transfers energy to a not yet observed component with interaction threshold in the knee region was proposed as the cause of the observed knee feature in the shower size spectrum \cite{t6,t7}. However, such a proposal has not received any support from the LHC experiment against the expectations. On the other hand Stenkin \cite{t8, t9} refuted the reality of the knee in the primary cosmic ray energy spectrum on the ground that the knee has been noticed observationally only in the electromagnetic component of EAS but not in the muonic and the hadronic components of EAS. In other words the knee feature in the primary cosmic ray energy spectrum is not consistently revealed from electromagnetic, muonic and hadronic components of EAS. Stenkin proposed an alternative explanation of the break in shower size spectrum in terms of coreless EAS \cite{t8, t9}. Further a new experiment PRISMA has been proposed to investigate the situation \cite{t10}.

While arguing against the astrophysical knee, Stenkin did not consider any effect of change in primary mass composition in the knee region on air shower muon and electron spectra \cite{t8}. Here it is worthwhile to mention that the almost all the well known models of the knee generally predict for a change in the mass composition of cosmic rays across the knee energy. For instances, the scenarios like rigidity dependent acceleration mechanism in the source or leakage from the Galaxy (which is also a rigidity dependent effect) predict for a heavier cosmic ray mass composition beyond the knee while the models based on nuclear photo-disintegration processes in the presence of a background of optical and soft UV photons in the source region predict for a lighter composition above the knee. The modern precise EAS experiments estimated primary energy spectra of different mass groups or even of various elements based on the deconvolution of either measured electron size distribution along with the information of muon content (as a function of electron size) or from a measured two-dimensional electron muon number distribution. Though conclusions of different experiments on primary mass composition in the knee region are not unequivocal, majority conclude that the knee represents the energy at which proton component exhibits cut-off \cite{t4} i.e. the knee of the spectrum has been ascribed as the proton knee. 

It is thus imperative to examine whether the primary knee feature is consistently revealed in electron and muon components of EAS when primary composition changes from lighter primaries to heavier primaries beyond the knee energy. This is precisely the objective of the present work. Our main emphasis will be to check whether the different EAS observables suggest for consistent spectral indices in the primary cosmic ray energy spectrum before and after the knee considering the fact that primary composition may or may not change across the knee. For this purpose we shall perform a detailed Monte Carlo simulation study of EAS using CORSIKA \cite{t11} in the concerned energy range and we will analyze different experimental data on size spectrum of various EAS observables to check the mutual consistency. We will also estimate the spectral indices of electron and muon size spectra for different primary composition scenario assuming primary cosmic ray energy spectrum has a knee. The hadronic component is not considered in this work as only few data in this regard are available and more importantly the uncertainties are quite large. 

The organization of the paper is as follows. In the next section the principle of deriving the cosmic ray energy spectrum from the EAS observables is outlined briefly in the framework of Bhabha-Heitler theory of electromagnetic cascade . In section III we describe our analysis of cosmic rsy EAS size spectra based on the Monte Carlo simulation study. The procedure adapted for the Monte Carlo simulation of cosmic ray EAS is discussed in the  subsection III-A. In the subsection III-B we evaluate spectral index of primary energy spectrum from the measured electron and muon size spectra considering different primary composition scenario. The expected shower size and muon size spectra for different mass composition scenario assuming the primary energy spectrum has a knee are obtained in the subsection III-C. Finally we discuss the findings and their probable explanations in the section IV.

\section{Primary energy spectrum from EAS observations and the knee}

Usually, cosmic ray EAS arrays employ scintillation detectors for detection of electrons, which is the dominating component among the charged particles in EAS. However, such detectors also detect other charged particles including  muons. So essentially EAS observations give information about total charged particle spectrum instead of electron size spectrum. The observational charged particle size (often known as shower size) spectrum in EAS is found to exhibit power law behavior i.e. 

\begin{equation}
\frac{dN}{dN_{ch}} \propto N_{ch}^{-\beta_{ch}}
\end{equation}
  

Though the development of EAS is a very complicated process that can be properly addressed only via Monte Carlo simulation technique but an idea of how electron and other secondary particle sizes are related to primary energy can be obtained based on the Bhabha - Heitler analytical approach of electromagnetic cascade \cite{t12, t13}. A cosmic ray particle interacts with the atmospheric nuclei while moving through the atmosphere and produced dominantly charged and neutral pions. There will be also secondary hadrons (leading particles). Neutral pions quickly decay to photons which subsequently initiate electromagnetic cascades. The charged pions may interact with atmospheric nuclei (thereby further produce secondary particles) or decay depending on their energy. The decay of charged pions yields muons and neutrinos. The energy dependence of total number electrons, muons and hadrons at shower maximum (at which the number of particles in a shower reaches its maximum) in EAS initiated by a nucleus with atomic mass number A and energy $E_{o}$ can be expressed as \cite{t12,t13} 

\begin{equation}
N_{i}^{max}= N^{o}_{i} E_{o}^{\alpha_{i}}
\end{equation}

where i stands for e (electron), $\mu$ (muon) and h (hadron). For pure electromagnetic cascade and under few simple approximations such as the all electrons have the same energy $E^{c}_{e}$ (which is the critical energy (85 MeV in air), at which ionization losses and radiative losses are equal) $\alpha_{e}$ is nearly equal to $1$. Similarly when all muons are considered to have the same energy $E^{c}_{\pi}$ (which is the energy at which the probability for a charged pion to decay and to interact are equal) and taking the charged pion production multiplicity is 10 (constant), $\alpha_{\mu} \sim 0.85$ \cite{t13}. When the effect of inelasticity is taken into consideration, $\alpha_{\mu}$ will be slightly higher, $\sim 0.90$ \cite{t13}. If one considers that total primary cosmic ray energy is distributed between electron and muon component, $\alpha_{e}$ will be slightly higher, about $1.05$ \cite{t13}.  

Two important points to be noted are (i) the total number of electrons increases with energy slightly faster than exactly linear whereas the total number of muons grows with energy slightly less than exactly linear. (ii) The electron number decreases with increasing mass number whereas muon number grows with mass number. 

After shower maximum, electron (and hadron) size decreases due to attenuation whereas muon size almost remain constant because of its large attenuation length. Hence at a observational level well passed the shower maximum, the equation (2) is not strictly valid, particularly for electrons and hadrons.

Assuming that the electron size spectrum and total charged particle size spectrum are more or less the same, from equations (1) and (2) one can infer the primary cosmic ray spectrum as follows

\begin{equation}
\frac{dN}{dE_{o}}=\frac{dN}{dN_{e}^{max}} \frac{dN_{e}^{max}}{dE_{o}} \propto E_{o}^{-\gamma}
 \end{equation}

where 
\begin{equation}
\gamma \equiv 1+\alpha_{e}(\beta_{e}-1) 
\end{equation}

will be the slope of primary cosmic ray differential energy spectrum. Since a sudden change in $\beta_{e}$ at a size corresponding to a primary energy of about 3 PeV is observed, consequently a change in $\gamma$ at 3 PeV is inferred which is the so called knee of the cosmic ray energy spectrum.      

Equations (2) and (3) imply that muon and hadron size spectra also should exhibit power law behavior with $\beta_{i} =1+(\gamma-1)/\alpha_{i}$. Since $\alpha_{\mu} < \alpha_{e}$, change in $\beta_{\mu}$ should be larger than $\beta_{e}$ for a change in $\gamma$. Observationally, however, no significant change in $\beta_{\mu}$ is found. This is why Stenkin objected the existence of a knee in the primary energy spectrum \cite{t8,t9}.              

Note that the semi-analytical expressions described above, though match reasonably well with the simulation results, are approximated description of cosmic ray cascade in the atmosphere. Moreover, the relation between electron size  and energy (Eq. 2) is valid only at shower maximum. So a detailed Monte Carlo simulation study needs to be done to draw any concrete conclusion in this regard.

\section{Monte Carlo simulation study of size spectrum}
 
In the present work we have simulated EAS for three different mass composition scenario: proton as primary over the whole energy range, secondly proton and Fe respectively as primary below and above the knee energy and finally Fe as primary over the whole energy range. Subsequently we explore whether a consistent mass composition scenario evolve from simultaneous study of electron and muon size spectra in the knee region. We evaluate $\alpha_{i}$ from simulation data for proton and iron primaries both below and above the knee and using the observed $\beta_{i}$ from experiments, we  subsequently estimate $\gamma$ following the equation (4) and check whether electron, muon and hadron observations give a consistent primary energy spectrum when primary composition is allowed to change across the knee. 

\subsection{Simulation procedure adopted}
The air shower simulation program CORSIKA (COsmic Ray SImulation for KAscade) (version 6.690) \cite{t11} is employed here for generating EAS events. The high energy (above $80 {\rm GeV/n}$) hadronic interaction model QGSJET 01 (version 1c) \cite{t14} has been used in combination with the low energy (below $80 {\rm GeV/n}$) hadronic interaction model UrQMD \cite{t15}. A relatively smaller sample has also been generated using the high-energy interaction model EPOS (version 2.1) \cite{t16}  and low energy interaction model GHEISHA (version 2002d) \cite{t17} to judge the influence of the hadronic interaction models on the results. Note that GHEISHA exhibits a few shortcomings \cite{t18,t19} but the low energy interaction models has no significant effect on the total number of secondary particles for primaries in the PeV energy range. 

The US-standard atmospheric model with planar approximation which works only for the zenith angle of the primary particles being less than $70^{\rm o}$ is adopted. The EAS events have been generated for proton and iron nuclei as primaries at several fixed energy points spreaded between $3 \times 10^{14}$ to $3 \times 10^{16}$ eV as well as over a continuous energy spectrum between $3 \times 10^{14}$ to $3 \times 10^{16}$ eV with differential energy spectrum slop -2.7and -3.1 below and above the knee ($3 \times 10^{15}$ eV) respectively.  The EAS events have been simulated at geographical positions correspond to experimental sites of KASCADE \cite{t20} and EAS-TOP \cite{t21}. The magnetic fields, observation levels, threshold energies of particle detection and zenith angles are provided accordingly.  

\subsection{Inferring Primary cosmic ray spectrum from measured EAS size spectra}
Only a few EAS experiments so far measured both $\beta_{ch}$ and $\beta_{\mu}$ before and after the knee. Here we would consider the results of two experiments, the KASCADE \cite{t22,t23} and EAS-TOP \cite{t24}. The KASCADE experiment was considered as one of the most precise air shower experiments in the world which was situated in the site of Forschungszentrum Karlsruhe (Germeny) at an altitude 110 m above sea level at 49.1$^o$ N, 8.4$^o$ E, covering an energy range from about 100 TeV to nearly 100 PeV and was in operation during October 1996 to 2003. The experiment consisted an array of electron and muon detectors, spread over 700 m$^2$ $\times$ 700 m$^2$, a central hadron calorimeter with substantial muon detection areas and a tunnel with streamer tube muon telescopes. This multi-detector system was used for the study of electromagnetic, muonic and hadronic components of EAS. The experiment was later extended to KASCADE-Grande in 2003 to study primary cosmic rays at higher energies. On the other hand the EAS-TOP array was located at Campo Imperatore, National Gran Sasso Laboratories in Italy, 2005 m a.s.l.,($820 \; g \;cm^{2}$) atmospheric depth. This multi-component experiment  consisted of detectors of the electromagnetic, muon, hadron and atmospheric Cherenkov light components for the study of EAS over the energy range 100 TeV to about 10 PeV. Two layers of streamer tubes with total surface area $12 \times 12 \; m^{2}$ was used for detection of EAS muons having threshold energy of 1 GeV.     
 
The results of these two experiments on $\beta_{ch}$ and $\beta_{\mu}$ are shown in Table 1. Note that the shower size ($N_{e}$) and muon size are generally evaluated from the experimental measured particle (electron/muon) densities by fitting with the lateral density distribution function. To minimize the bias by the functional form of the muon lateral distribution function, KASCADE experiment introduced the quantity truncated muon number which is essentially the muon size within 40 m and 200 m core distance.   

\begin{table*}[ht]
\begin{center}
\caption{The measured spectral indices of primary energy spectrum below and above the knee from the electron and the muon size spectra of KASCADE and EAS-TOP observations}
\begin{tabular}  
{|c|c|c|c|} \hline 

  Experiment & Component & $\beta_{<knee}$      & $\beta_{> knee}$   \\   \hline
   KASCADE   & charged particles  & $2.45 \pm 0.06$      & $2.94 \pm 0.12$   \\   \hline
	 KASCADE   &  muon ($>490$ MeV)    & $3.05 \pm 0.006$     & $3.27 \pm 0.01$   \\   \hline
   EAS-TOP    & charged particles  & $2.61 \pm 0.01$      & $3.01 \pm 0.06$   \\   \hline
   EAS-TOP    & muon ($> 1$ GeV)     & $3.12 \pm 0.03$      & $3.67 \pm 0.07$   \\   \hline
   
\end{tabular}
 
\end{center}
\end{table*}

Using the public data of KASCADE experiment provided through KCDC \cite{t25} we estimated $\beta$ ourselve. For vertical air showers ($\theta < 18^{o}$), we find $\beta$ equals to $2.54 \pm 0.06$ and $2.97 \pm 0.05$ below and above the knee are respectively for total charged particles and $2.96 \pm 0.08$ and $3.24 \pm 0.06$ for muons below and above the knee respectively which are closed to the KASCADE reported $\beta$.

To estimate $\alpha$ we exploit Monte Carlo simulation method. The figure 1(a) displays the variation of total charged particle number in EAS obtained with Monte Carlo simulation as a function of energy at KASCADE location for proton primary whereas the variation of muon content with primary energy in proton induced EAS is shown in figure 1(b). Power law fits to the data points are also shown in both the figures. We find that the dependence of shower size on primary energy can be described by a power law with constant spectral index as given in equation (2). We have also checked whether the data suggest different spectral slops at lower and higher energies by fitting the data below and above the knee separately. But the so fitted slops are found only to differ within the error limits of the single constant spectral index. The estimated power law indices ($\alpha_{ch}$ and $\alpha_{\mu}$) are displayed in table 1 for proton primary. In figure 2 we have plotted the electron and muon sizes in Fe initiated EAS as a function of primary energy. The $\alpha_{ch}$ and $\alpha_{\mu}$ for Fe primary are also evaluated from power law fitting and are shown in table 2.      

\begin{figure*}[ht]
\renewcommand{\thesubfigure}{\relax}
\subfigure[]
{\includegraphics[scale=0.4]{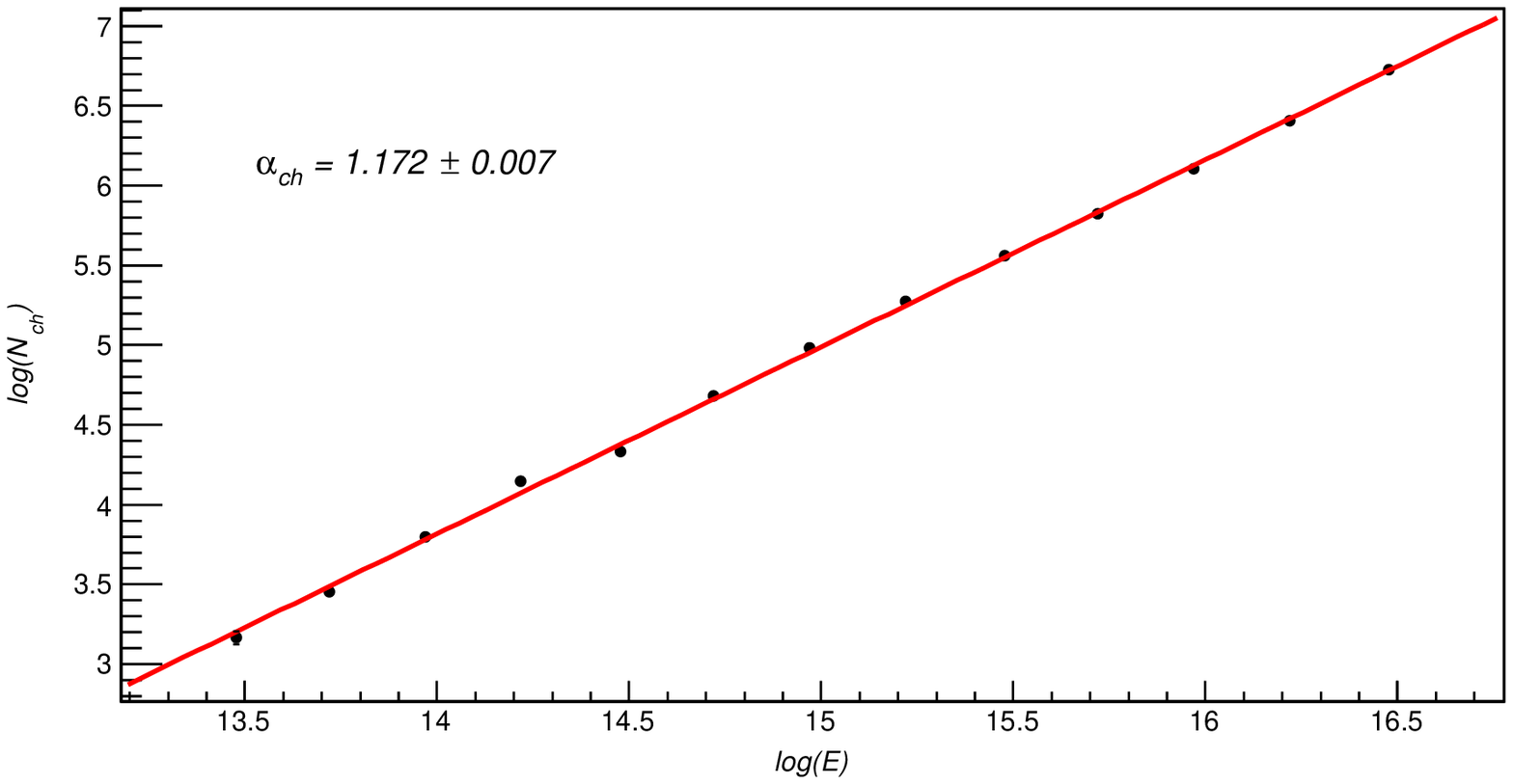} }
\renewcommand{\thesubfigure}{\relax}
\subfigure[]
{\includegraphics[scale=0.4]{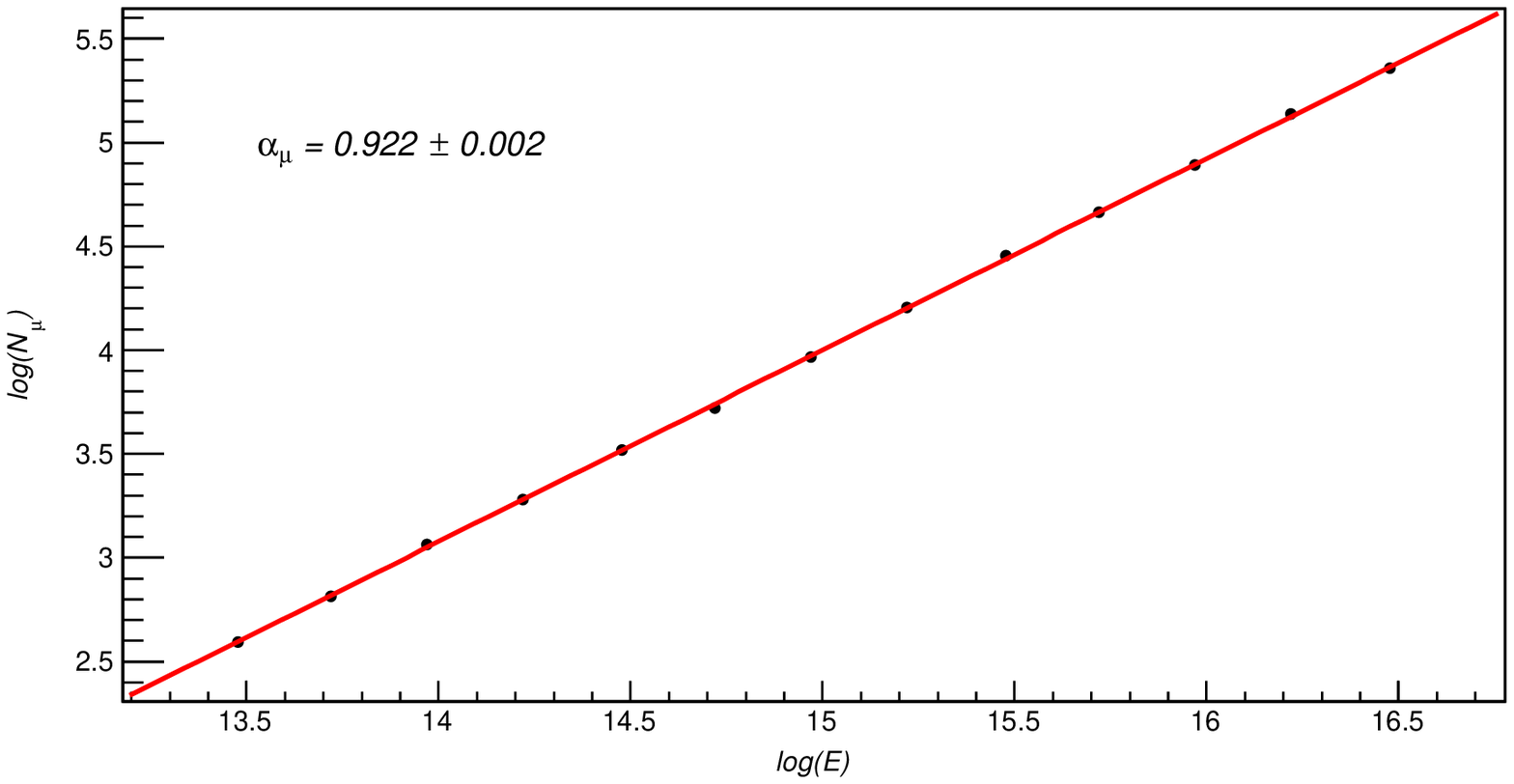}}
\caption{(Color online) Energy dependence of (a) total charged particles and (b) muon content in proton induced EAS at KASCADE location from the Monte Carlo simulation data.}
\end{figure*}

\begin{figure*}[ht]
\renewcommand{\thesubfigure}{\relax}
\subfigure[]
{\includegraphics[scale=0.4]{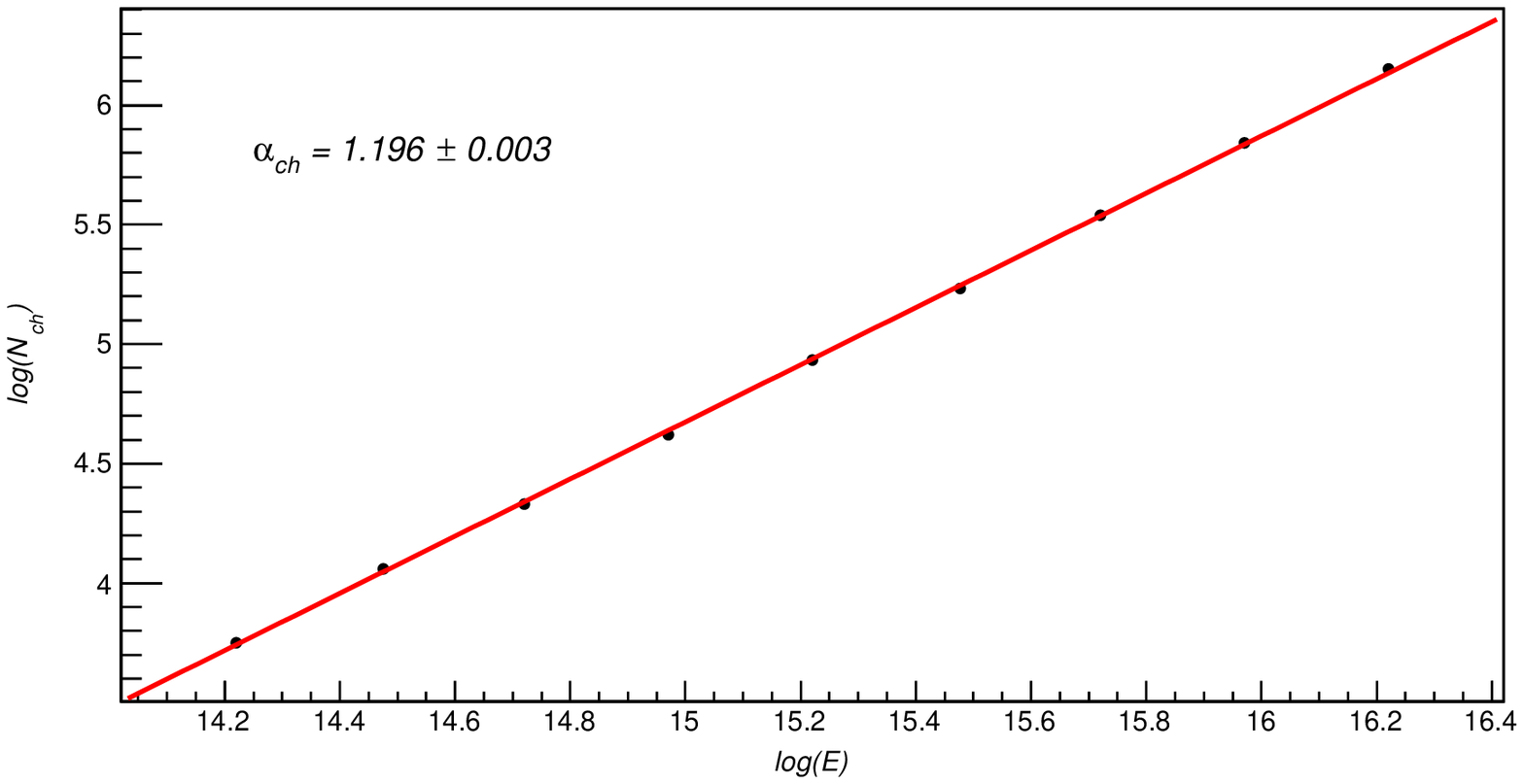} }
\renewcommand{\thesubfigure}{\relax}
\subfigure[]
{\includegraphics[scale=0.4]{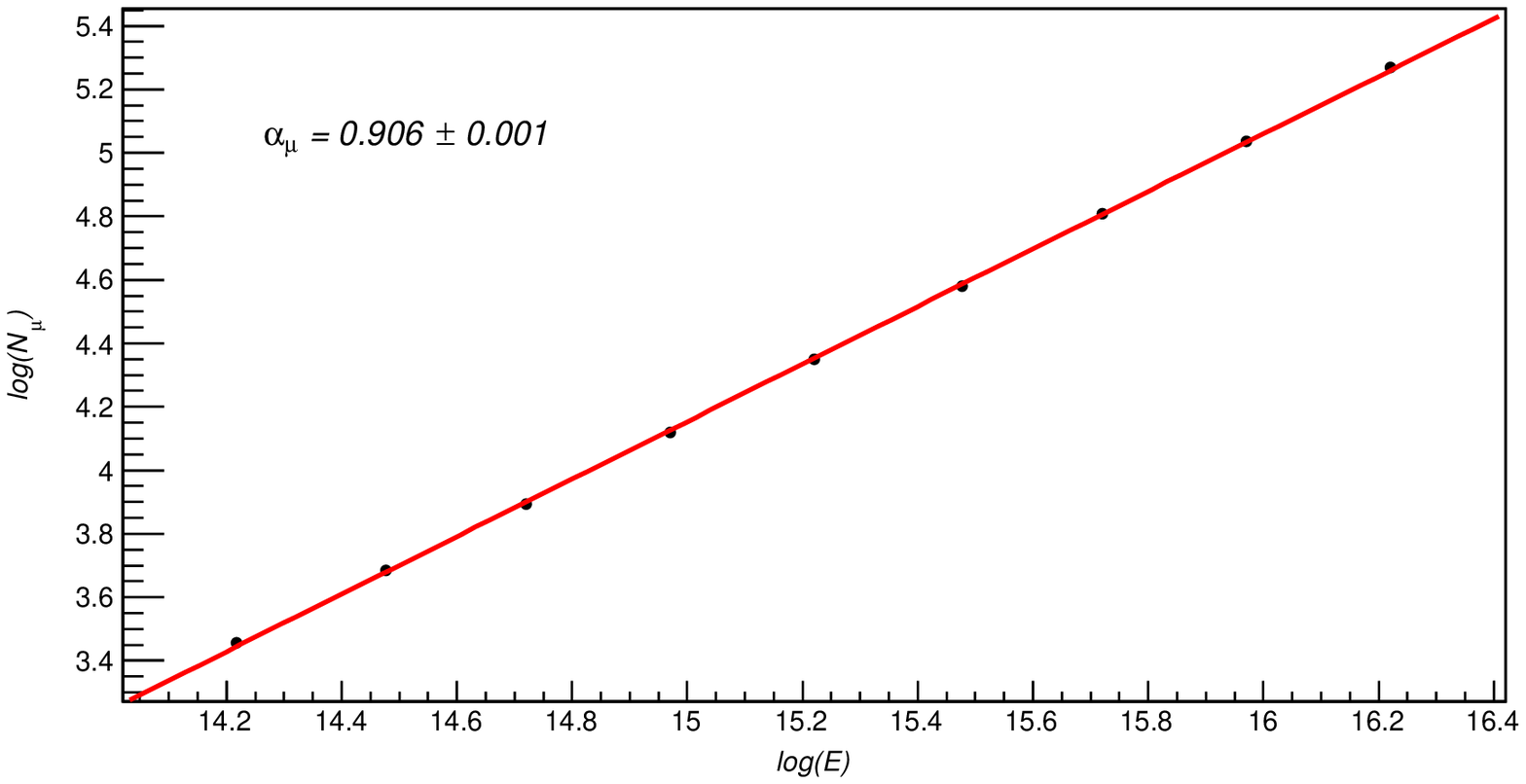}}
\caption{(Color online) Same as Figure 1 but in Fe initiated EAS.}
\end{figure*}

Since $\beta_{i}$s are known from observations, we have estimated $\gamma$ straightway using the expression (4). We considered both proton and Fe as primaries below the knee as well as above the knee and evaluate $\gamma$. Subsequently we compute $\delta \gamma$ across the knee. The results are given in Table 1 for the KASCADE measurements. It is noticed that no consistent $\gamma$s below and above the knee emerge from the KASCADE measured electron and muon spectra irrespective of the primary composition. The $\delta \gamma$s from the observed electron and muon spectra also differ significantly.



\begin{table*}[ht]
\begin{center}
\caption{Spectral indices of primary energy spectrum below and above the knee from the electron and the muon size spectra of KASCADE observations}
\begin{tabular}  
{|c|c|c|c|c|c|c|c|} \hline 

Primary & Primary & Secondary & $\alpha_{<knee}$ & $\alpha_{>knee}$ & $\gamma_{<knee}$ & $\gamma_{>knee}$ & $\Delta \gamma$ \\

before the knee & after the knee&  & & & & &  \\ \hline
Proton &  Proton   &  electron    & $1.172 \pm 0.007$ & $1.172 \pm 0.007$ & $2.70 \pm 0.08 $ & $3.27 \pm 0.16$ & $0.57\pm 0.24 $     \\
       &    &  Muon       & $0.922 \pm 0.002$ & $0.922 \pm 0.002$ &  $2.89 \pm 0.01$ &$3.09 \pm 0.02$ & $0.20 \pm 0.04$     \\  
                            &    &  ($>490$ MeV) & & & & & \\ \hline

Proton &  Fe  &  charged particles    & $1.172 \pm 0.007$ & $1.196 \pm 0.003$ & $2.70 \pm 0.08$ & $3.32 \pm 0.14$ & $0.62 \pm 0.22$     \\
       &    &  Muon     & $0.922 \pm 0.002$ & $0.906 \pm 0.001$ & $2.89 \pm 0.01$& $3.05 \pm 0.02$ & $0.16\pm 0.03$     \\  
                            &    &  ($>490$ MeV) & & & & & \\ \hline

Fe &  Fe  &  charged particles    & $1.196 \pm 0.003$ & $1.196 \pm 0.003$ & $2.73 \pm 0.08$ & $3.32 \pm 0.14$ & $0.59 \pm 0.22$     \\
       &    &  Muon     & $0.906 \pm 0.001$ & $0.906 \pm 0.001$ &  $2.86 \pm 0.01$ & $3.05 \pm 0.02$ & $0.19\pm 0.03$     \\  
                           &    &  ($>490$ MeV) & & & & & \\ \hline
\end{tabular}
 
\end{center}
\end{table*}

Results of a similar analysis for the EAS-TOP electron and muon spectra are displayed in figures 3 nad 4 from simulation data and in Table 2. In EAS-TOP location the $\alpha$ of charged particles for proton primary is found quite small than that for the KASCADE location which suggests that $\alpha$ changes with atmospheric depth and approaches to one at shower maximum as predicted by the cascade theory. For Fe primary, however, no significant difference in $\alpha$ of charged particles noticed in two stated locations. This is probably due to the fact that air showers reaches to its maximum development much earlier for heavier primaries, so even at EAS-TOP altitude, PeV energy Fe initiated showers are quite old. The spectral index (of primary cosmic ray energy spectrum) derived separately from the EAS-TOP observed electron and muon size spectra is found somewhat mutually consistent when cosmic ray primary is dominantly Fe, both before and after the knee. The $\delta \gamma$s from the observed electron and muon spectra also found mutually consistent for unchanging Fe dominated primary.      

\begin{figure*}[ht]
\renewcommand{\thesubfigure}{\relax}
\subfigure[]
{\includegraphics[scale=0.4]{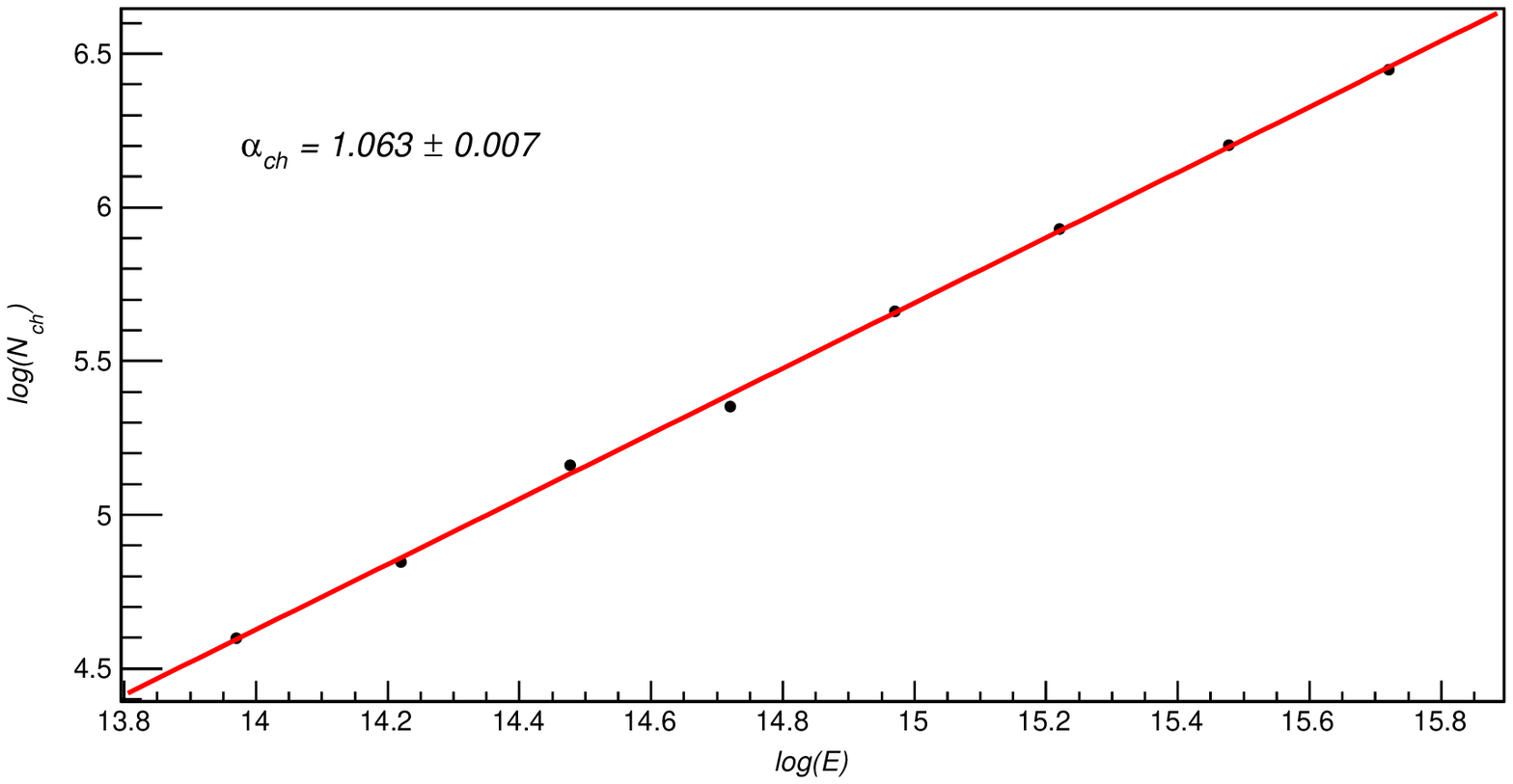} }
\renewcommand{\thesubfigure}{\relax}
\subfigure[]
{\includegraphics[scale=0.4]{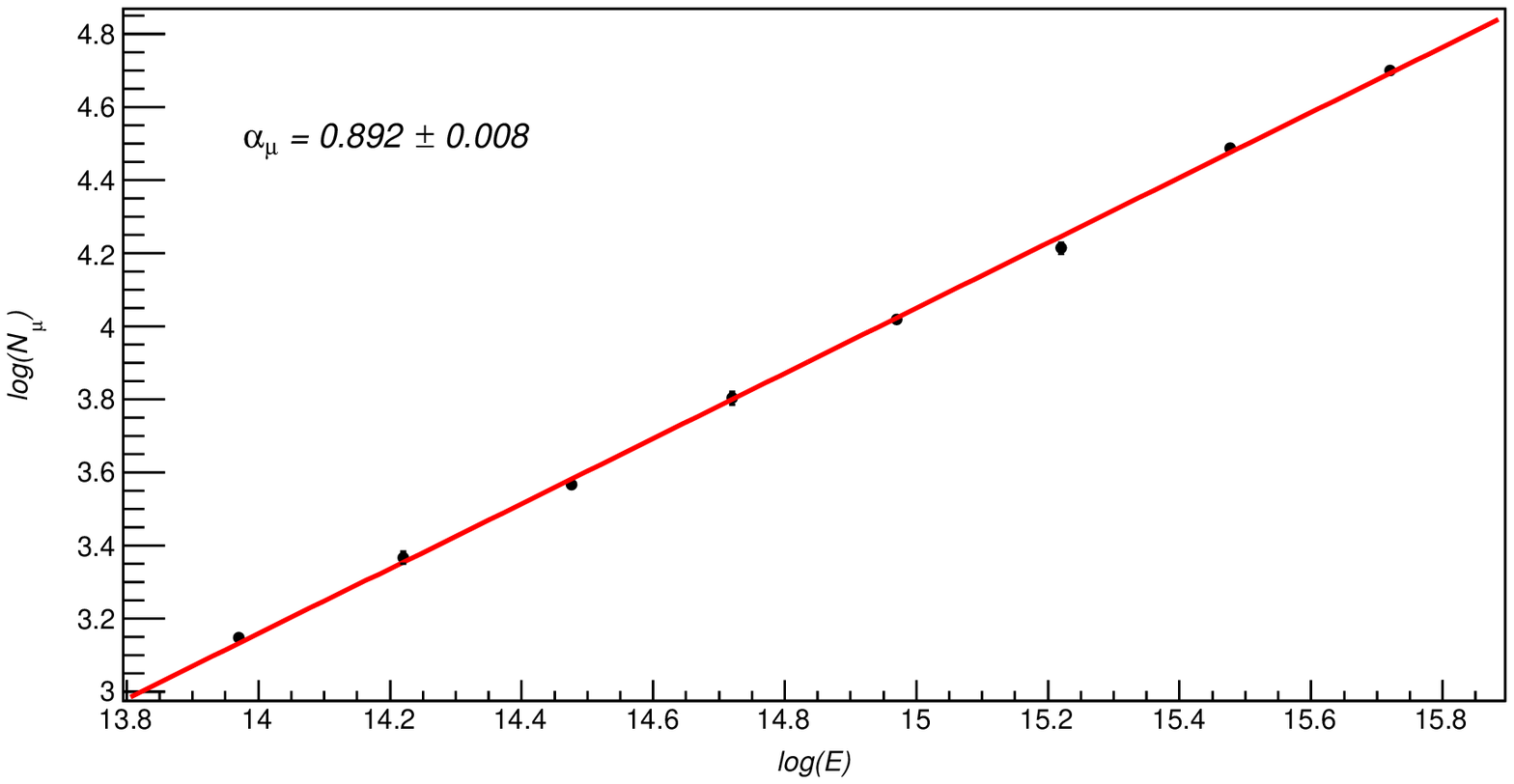}}
\caption{(Color online) Energy dependence of (a) total charged particles and (b) muon content in proton induced EAS at EASTOP location.}
\end{figure*}

\begin{figure*}[ht]
\renewcommand{\thesubfigure}{\relax}
\subfigure[]
{\includegraphics[scale=0.4]{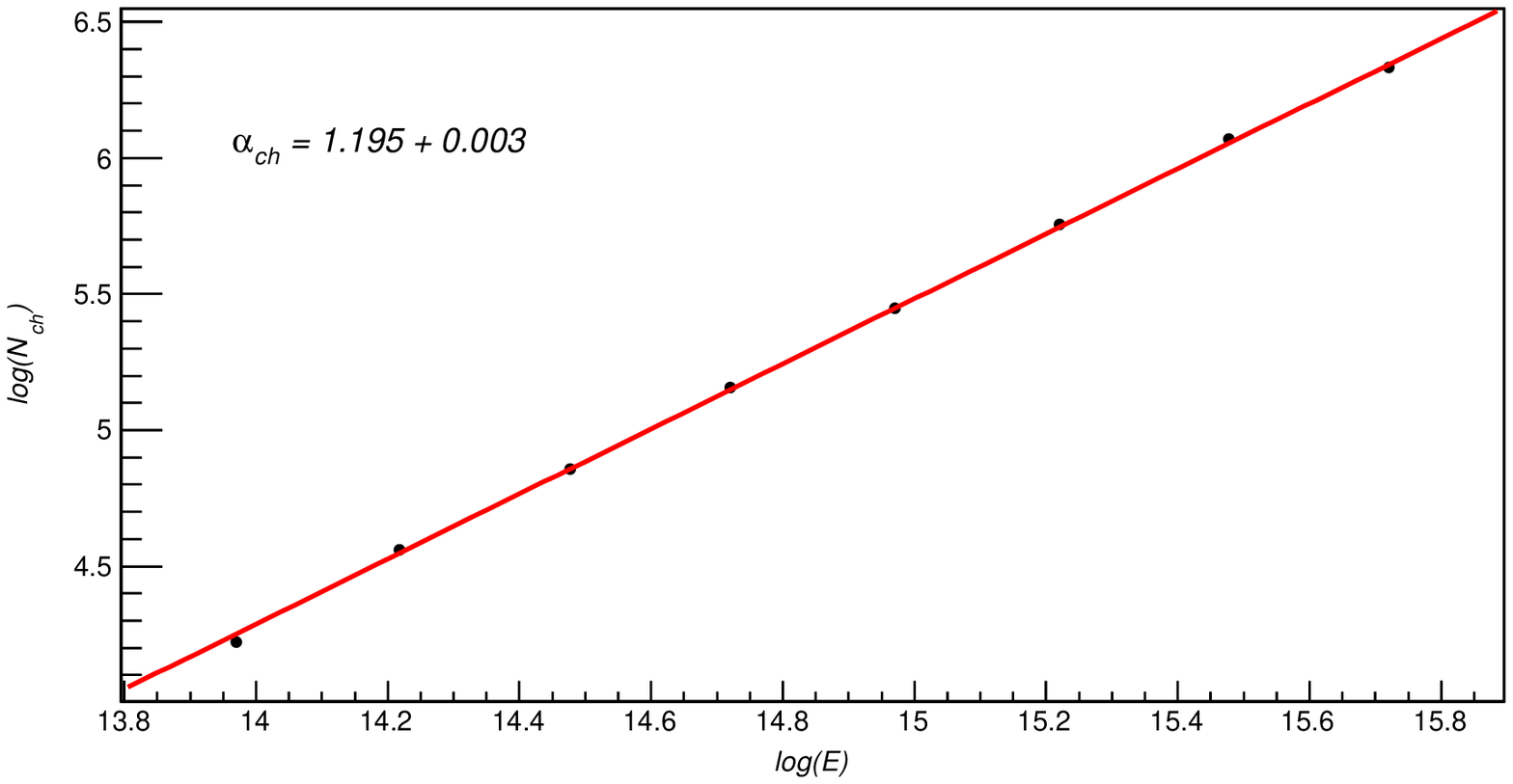} }
\renewcommand{\thesubfigure}{\relax}
\subfigure[]
{\includegraphics[scale=0.4]{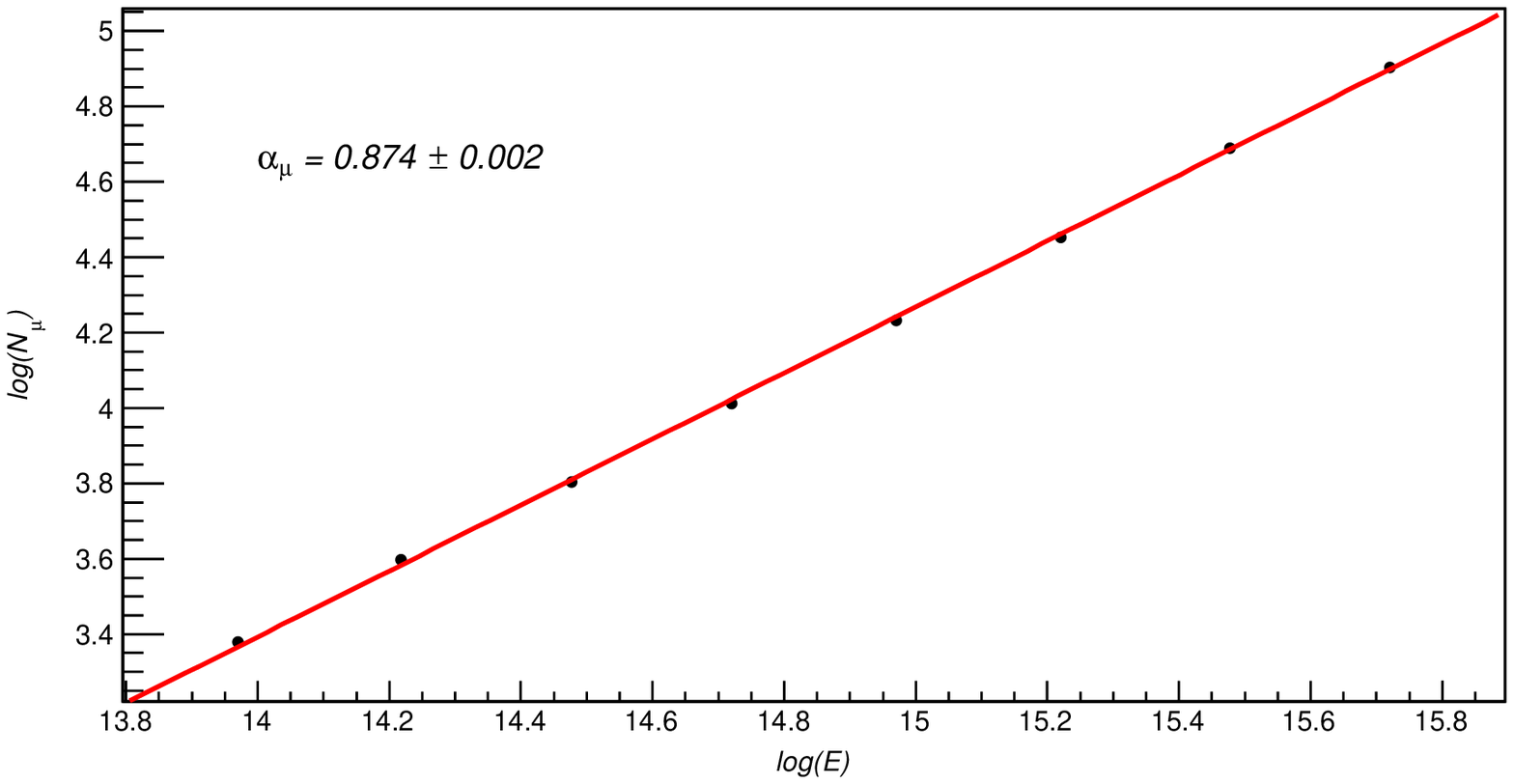}}
\caption{(Color online) Same as Figure 3 but in Fe initiated EAS .}
\end{figure*}

\begin{table*}[ht]
\begin{center}
\caption{Spectral indices of primary energy spectrum below and above the knee from the electron and the muon size spectra of EAS-TOP observations}
\begin{tabular}  
{|c|c|c|c|c|c|c|c|} \hline 

Primary & Primary & Secondary  & $\alpha_{<knee}$ &$\alpha_{>knee}$ & $\gamma_{<knee}$ & $\gamma_{>knee}$ & $\Delta \gamma$ \\

before the knee & after the knee & & & & & &  \\ \hline
Proton &  Proton   &  charged particles    & $1.063 \pm 0.007$ & $1.063 \pm 0.007$ &  $2.71 \pm 0.02$ & $3.14 \pm 0.07$ & $0.43\pm 0.09$     \\ 
       &    &  Muon        & $0.892 \pm 0.008$ & $0.892 \pm 0.008$ & $2.89 \pm 0.04$ & $3.38 \pm 0.09$ & $0.49\pm 0.13$     \\  \hline
 Proton & Iron     &  charged particles    & $1.063 \pm 0.007$ & $1.195 \pm 0.003$ & $2.71 \pm 0.02$ & $3.40 \pm 0.08$ & $0.69\pm 0.10$     \\  
       &    &  Muon        & $0.892 \pm 0.02$ &  $0.874 \pm 0.002$ & $2.89 \pm 0.04$ & $3.33 \pm 0.07$ & $0.44\pm 0.11$     \\ \hline 
Iron & Iron     &  charged particles    & $1.195 \pm 0.003$ & $1.195 \pm 0.003$ & $2.92 \pm 0.02$ & $3.40 \pm 0.08$ & $0.48\pm 0.10$     \\  
       &    &  Muon        & $0.874 \pm 0.002$ &  $0.874 \pm 0.002$ & $2.85 \pm 0.03$ & $3.33 \pm 0.07$ & $0.48\pm 0.10$     \\ \hline 

\end{tabular}
 
\end{center}
\end{table*}

\subsection{Electron and muon spectra for astrophysical knee}
It appears that the main difficulty of arriving a consistent knee from simultaneous charged particles and muon spectra in EAS from the KASCADE experiment is the very small spectral slope difference in muon spectrum ($\Delta \beta_{mu}$) across the knee relative to the spectral slope difference in charged particle spectrum ($\Delta \beta_{ch}$). Here we shall follow a reverse process, we shall estimate the expected spectral slopes in charged particle and muon spectra for different primary composition scenario assuming that the primary energy spectrum has a knee. The spectral index of the primary energy spectrum below the energy 3 PeV is taken as $-2.7$ whereas above 3 PeV it is assumed as $-3.1$. The EAS are generated from the minimum energy of 100 TeV and only vertical showers ($Z <18^{o}$) are generated.    

The charged particle and muon size spectra at KASCADE location from the simulation results are displayed in figures 5. We considered unchanged proton and Fe mass compositiona over the entire energy range as well as a change in mass composition after the knee from pure proton to pure iron. The knee structure is found present in both electron and muon size spectra for all the mass composition scenario considered. The $\beta$ value obtained from the simulation results are displayed in Table 4 for the different composition scenario. To estimate the $\beta$ values in electron and muon size spectra we multiply the differential total charged particle (muon) numbers by some suitable power (selected by varying the power index slowly) of total charges particles (muons) to emphasize the small difference in slope and plot it against the total charged particles (muons) in log-log scale. It is found that the points below and above a certain total charged particle number have  distinct slopes. The best fitted slopes give the $\beta$ below and above the size knee whereas the  crossing point of the two straight lines (in log-log scale) give the position of the knee in the size spectra.     

The spectral index of total charged particle spectrum above the knee obtained from the simulation results is found slightly lower than the observational result whereas for muon spectrum the spectral index below the knee from the simulation data is found slightly larger than the observations which is of not much importance as we assume spectral index of primary spectrum arbitrarily. The spectral indices for proton and iron primaries are found close. When composition changes across the knee it is noticed that the spectral index below (or above) the knee depends not only primary composition below (above) the knee of the primary energy spectrum but also the composition above (below) the knee of the energy spectrum, unless points close to the knee in the size spectra are excluded to determine the spectral index. There are few other noteworthy points : \\

i) the position of the knee in the charged particles and muon spectra also influence by the primary composition both below and above the knee of the cosmic ray energy spectrum, 

ii) the knee in the muon spectrum is slightly more revealing in comparison to that in the electron spectrum for pure proton or Fe primaries over the entire energy range but the same may not be true when primary composition changes across the knee, 

and iii) for proton primary before the knee and Fe primary after the knee the muon spectrum exhibits a break not only in the spectral index but also in the flux. The later feature is due to larger muon size in Fe initiated EAS in comparison to proton induced EAS. 

\begin{figure*}[ht]
\renewcommand{\thesubfigure}{\relax}
\subfigure[]
{\includegraphics[scale=0.4]{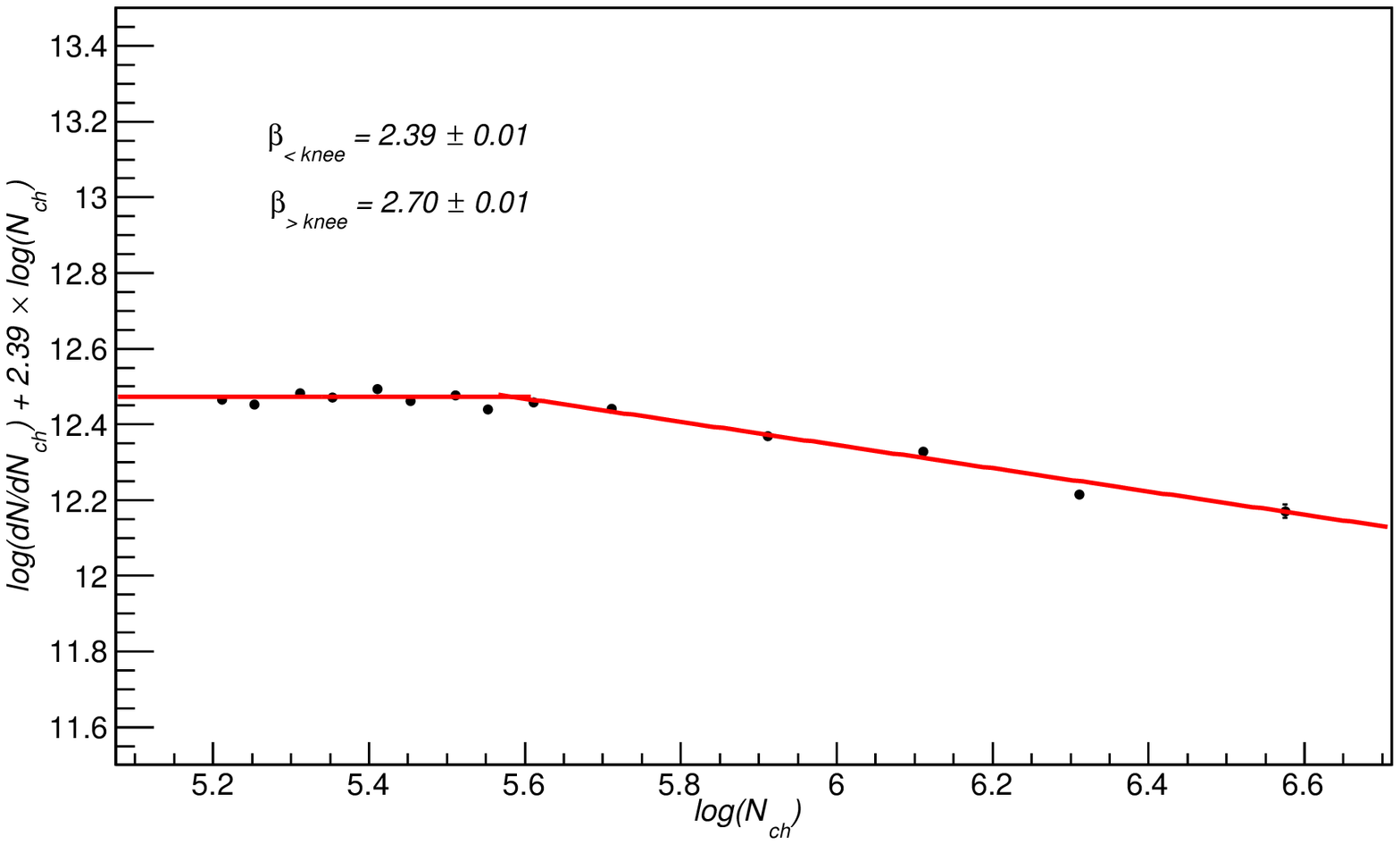} }
\renewcommand{\thesubfigure}{\relax}
\subfigure[]
{\includegraphics[scale=0.4]{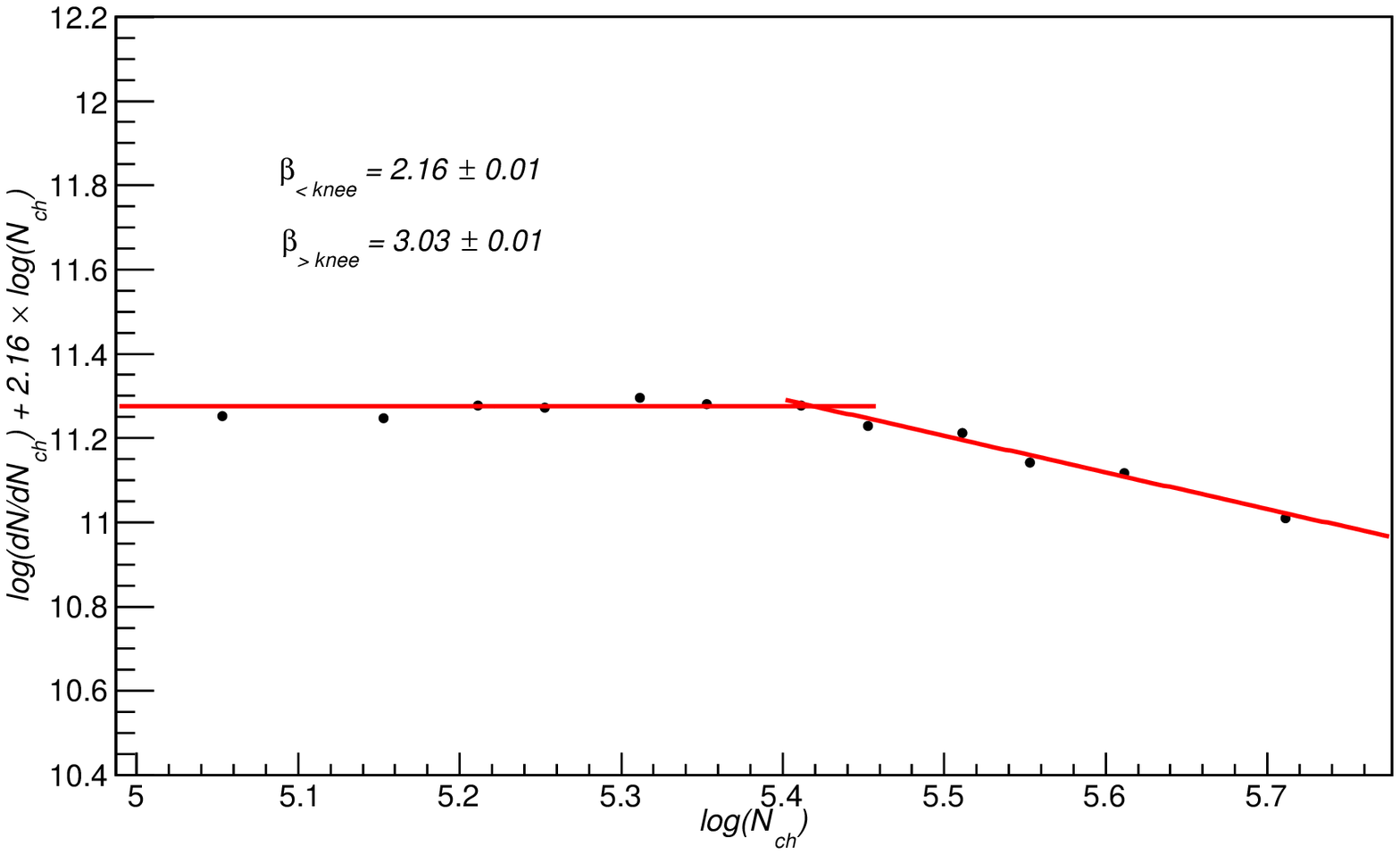}}
\subfigure[]
{\includegraphics[scale=0.4]{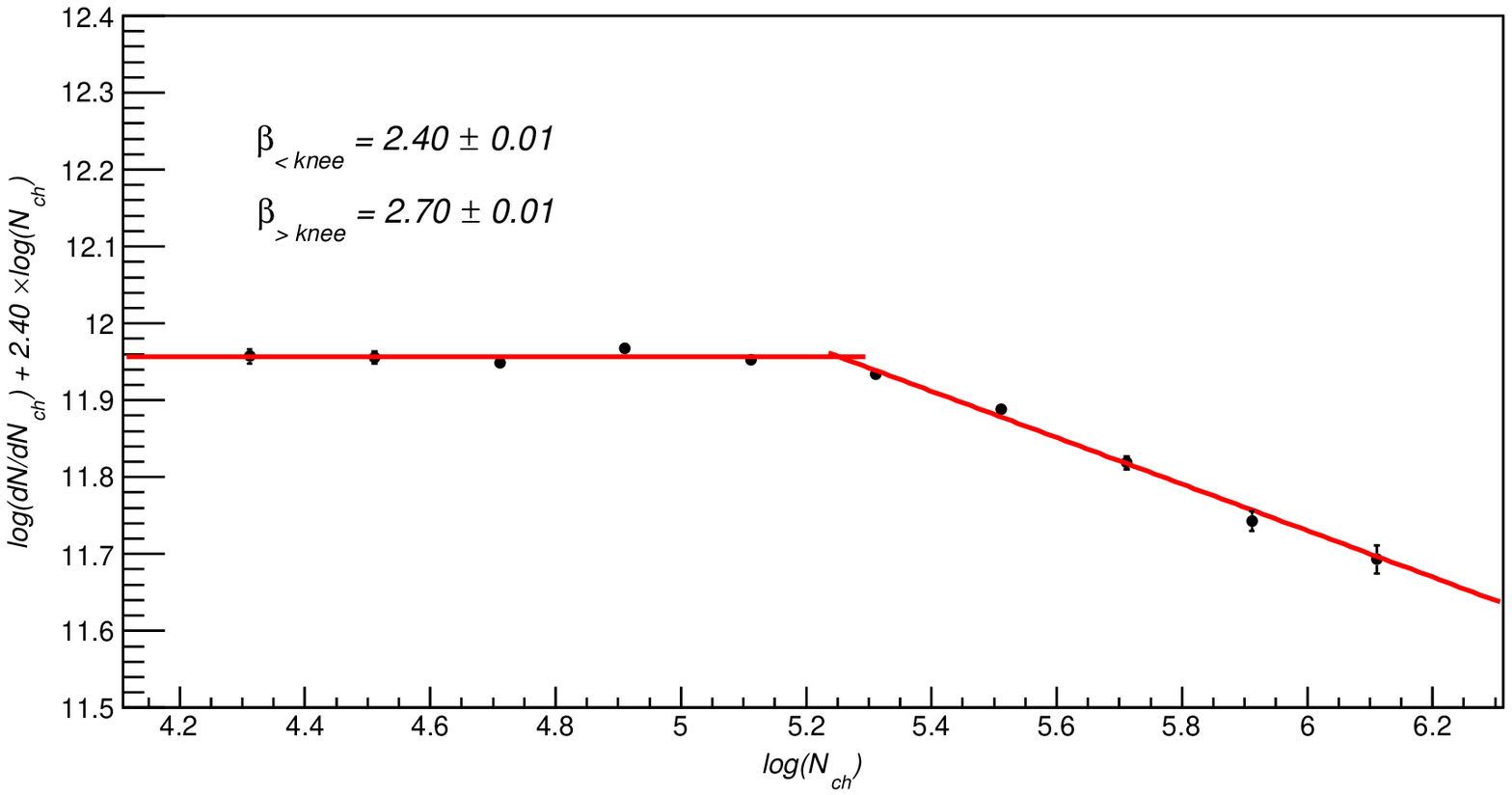}}
\caption{(Color online) Expected total charged particle size spectrum for different mass composition scenario across the knee (a) unchanged proton primary (b) proton below the knee and Fe above the knee and (c) unchanged Fe primary.}
\end{figure*}

\begin{figure*}[ht]
\renewcommand{\thesubfigure}{\relax}
\subfigure[]
{\includegraphics[scale=0.4]{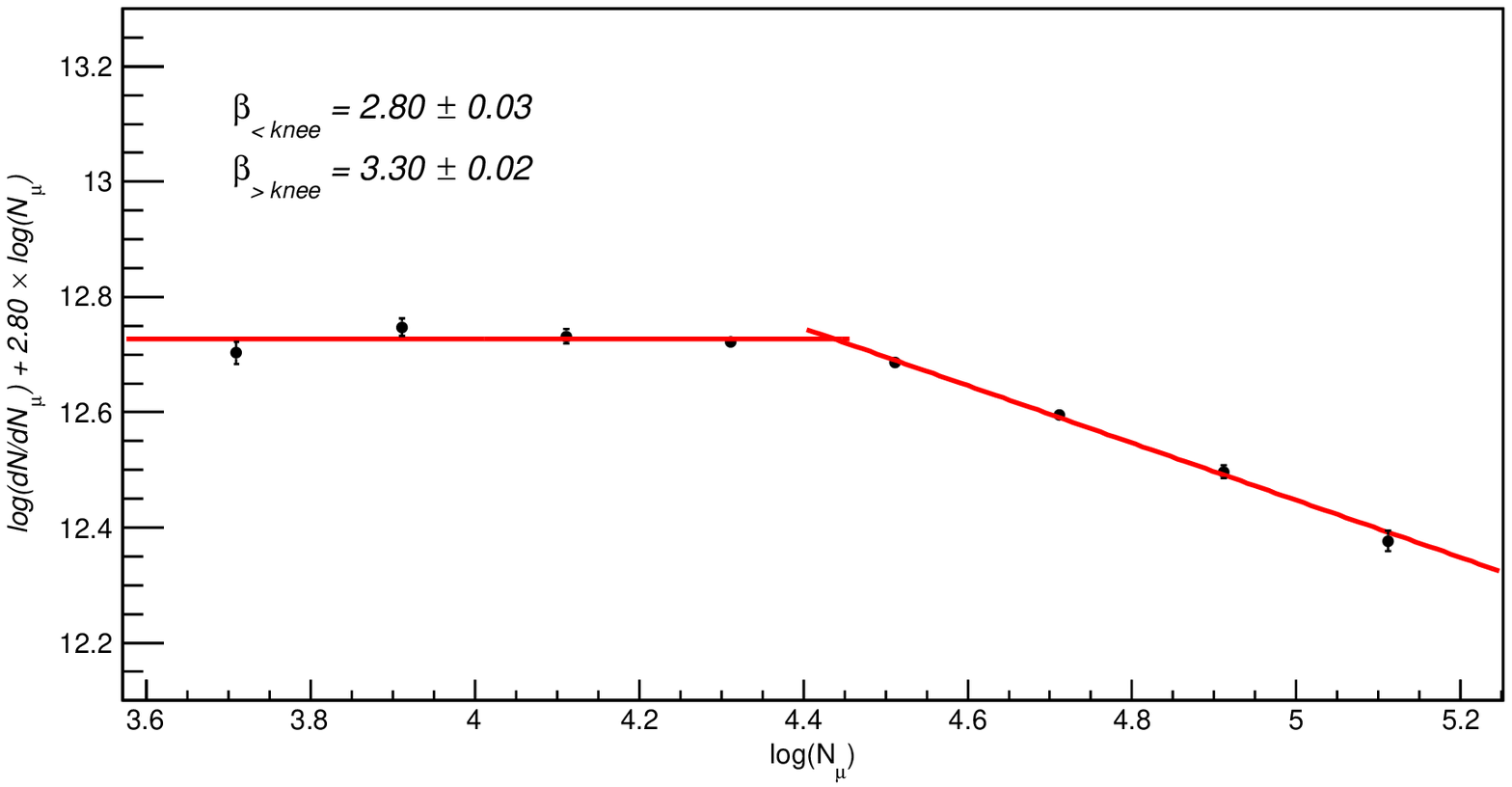} }
\renewcommand{\thesubfigure}{\relax}
\subfigure[]
{\includegraphics[scale=0.4]{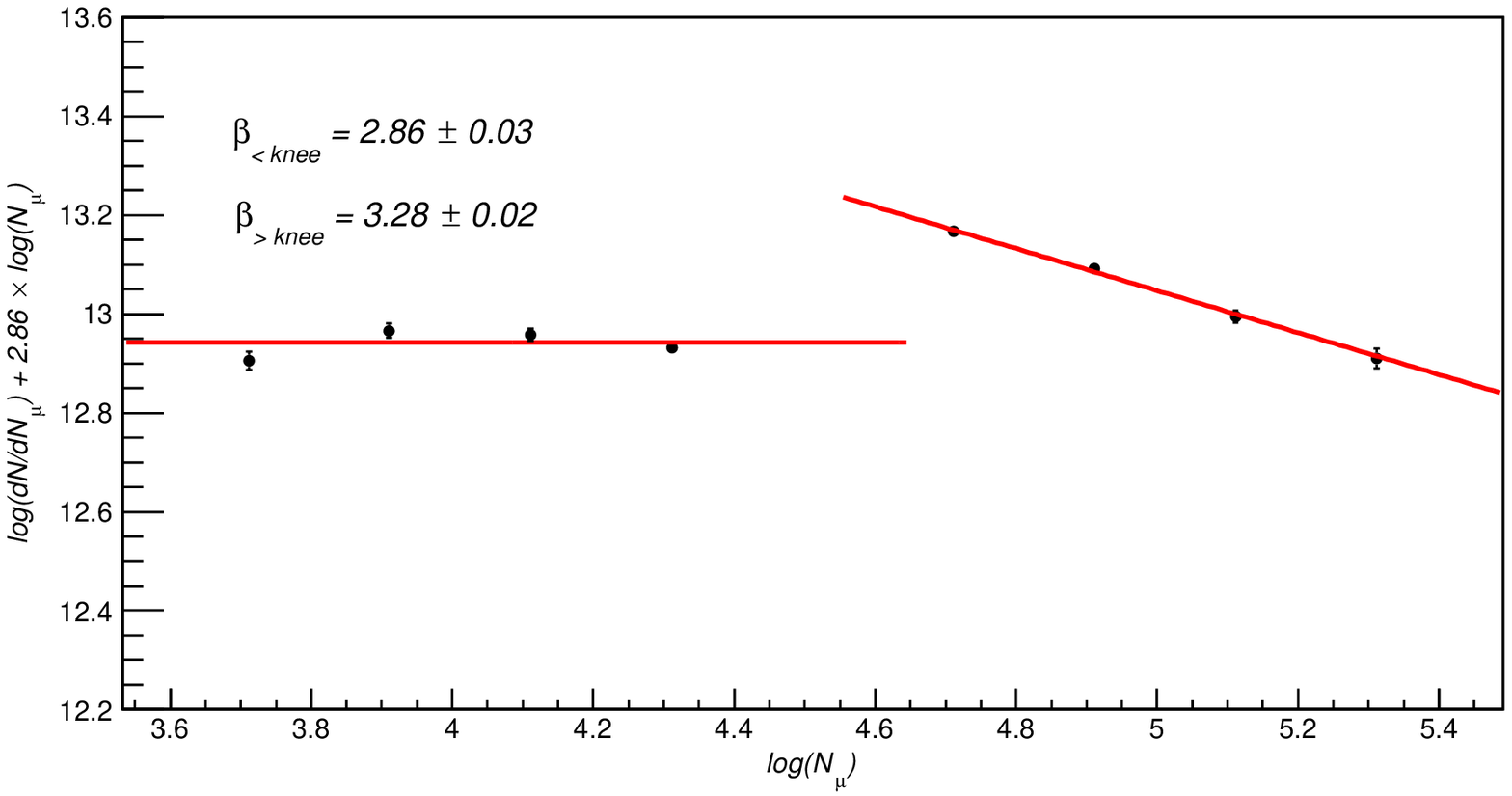}}
\subfigure[]
{\includegraphics[scale=0.4]{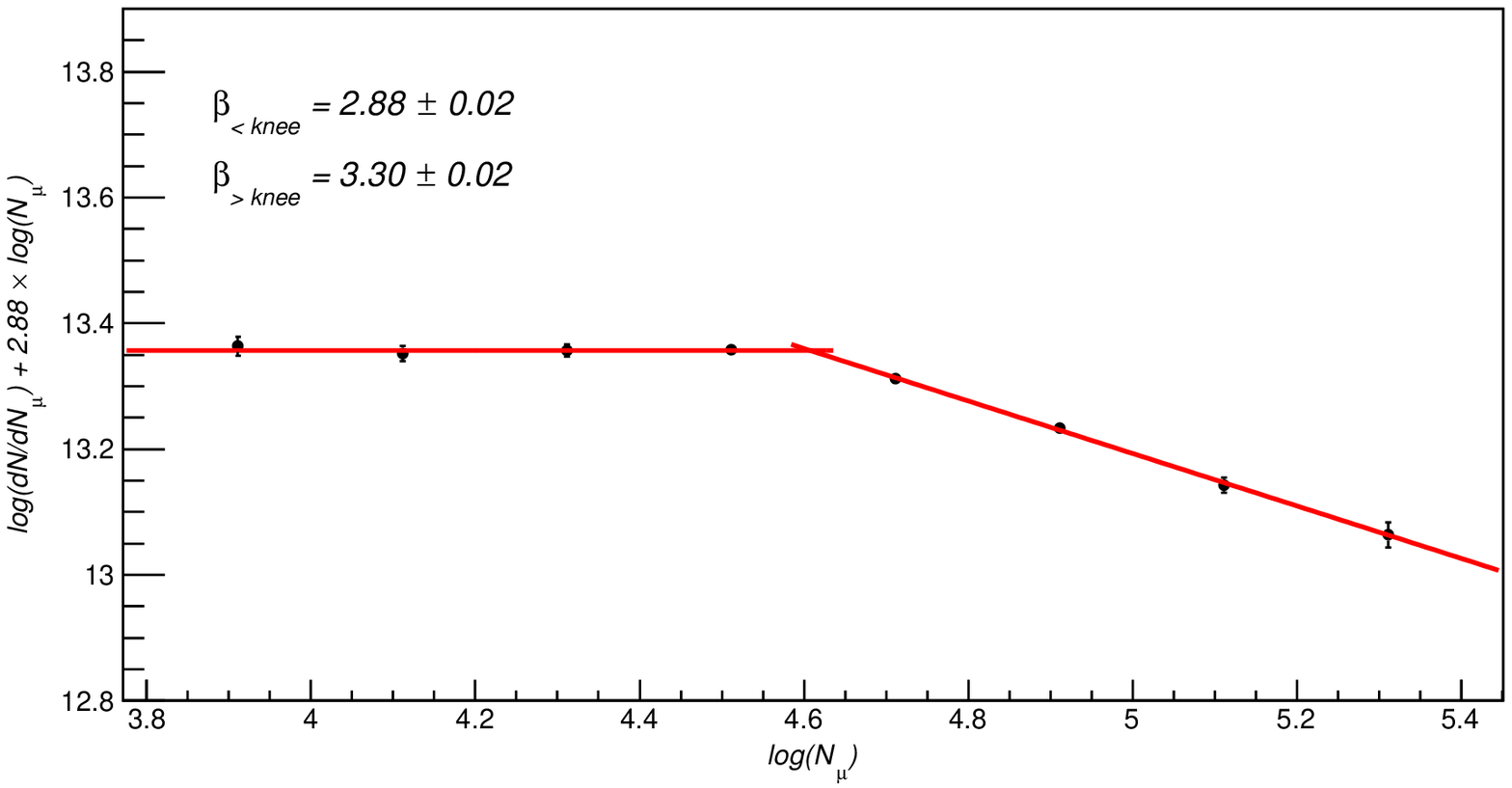}}
\caption{(Color online) Same as figure 5 but for muon spectrum}
\end{figure*}

\begin{table*}[ht]
\begin{center}
\caption{Spectral indices of the simulated charged particles and the muon size spectra for cosmic ray energy spectrum with the knee}
\begin{tabular}  
{|c|c|c|c|c|c|} \hline 

 Primary & Primary & Secondary  & $\beta_{<knee}$     & $\beta_{> knee}$  & $\Delta \beta$  \\

before the knee & after the knee  & & & & \\ \hline
Proton &  Proton   &  charged particles    & $2.39 \pm 0.01$ & $2.70 \pm 0.01$ & $0.31 \pm 0.02$ \\ 
       &    &  Muon        & $2.80 \pm 0.03$ & $3.30 \pm 0.02$ & $0.50 \pm 0.05$ \\  \hline
 Proton & Iron     &  charged particles    & $2.16 \pm 0.01$ & $3.03 \pm 0.01$ & $ 0.87 \pm 0.02$ \\  
  & &  Muon        &  $2.86 \pm 0.03$ &  $3.28 \pm 0.02$ & $0.42 \pm 0.05$  \\ \hline 
Iron & Iron     &  charged particles    & $2.40 \pm 0.01$ &  $2.70 \pm 0.01$ & $0.30 \pm 0.02$ \\  
       &    &  Muon        & $2.88 \pm 0.02$ &  $3.30 \pm 0.02$ & $0.42 \pm 0.04$  \\ \hline


\end{tabular}
 
\end{center}
\end{table*}

The modern EAS experiment usually employ two-dimensinal plots of total charged particle and muon size spectra to evaluate primary energy spectrum and composition. From simulation data two-dimensional plots of total charged particles and muon size spectra for different composition scenario are also obtained and depicted in figures 6 at KASCADE location. An interesting observation is that the knee is not clearly visualized from the two-dimensional plots. Since Fe induced EAS contains lower electrons and higher muons in compare to proton induced EAS, the two dimensional figure exhibits some mismatch in shower and muon sizes around the knee for a sharp change in composition from proton to Fe across the knee which is not observed experimentally.    
   
\begin{figure*}[ht]
\renewcommand{\thesubfigure}{\relax}
\subfigure[]
{\includegraphics[scale=0.6]{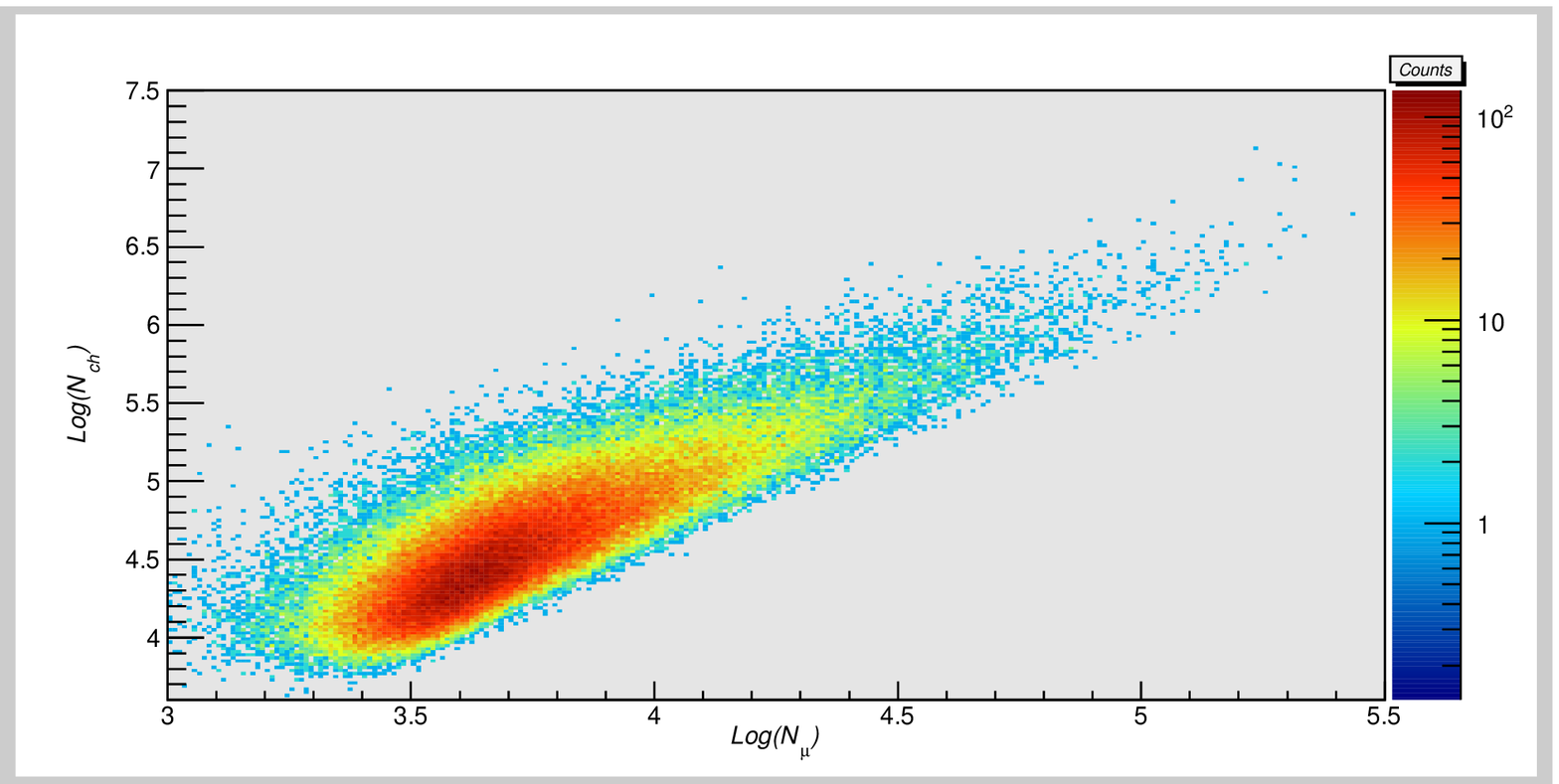} }
\renewcommand{\thesubfigure}{\relax}
\subfigure[]
{\includegraphics[scale=0.6]{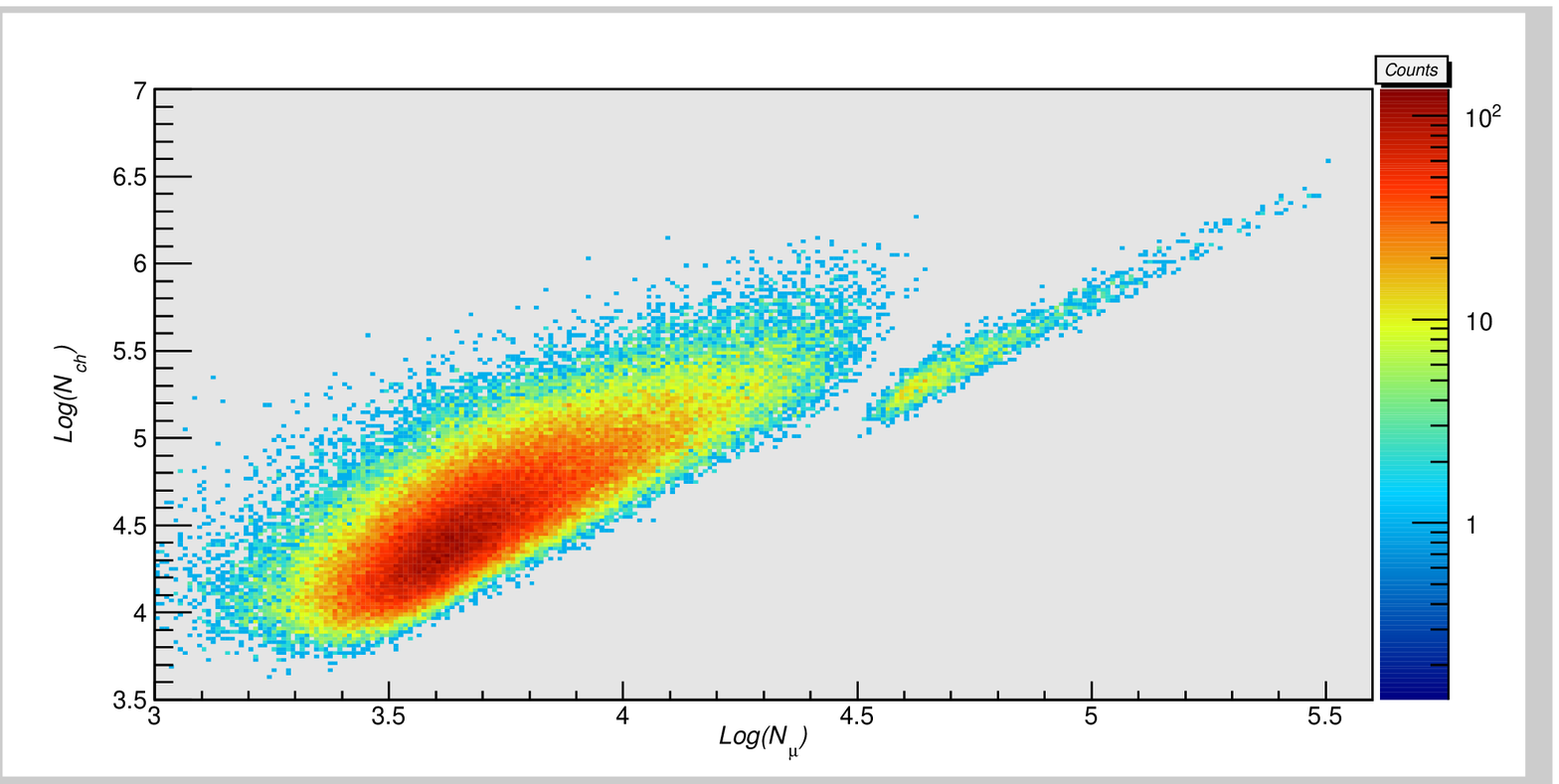}}
\subfigure[]
{\includegraphics[scale=0.6]{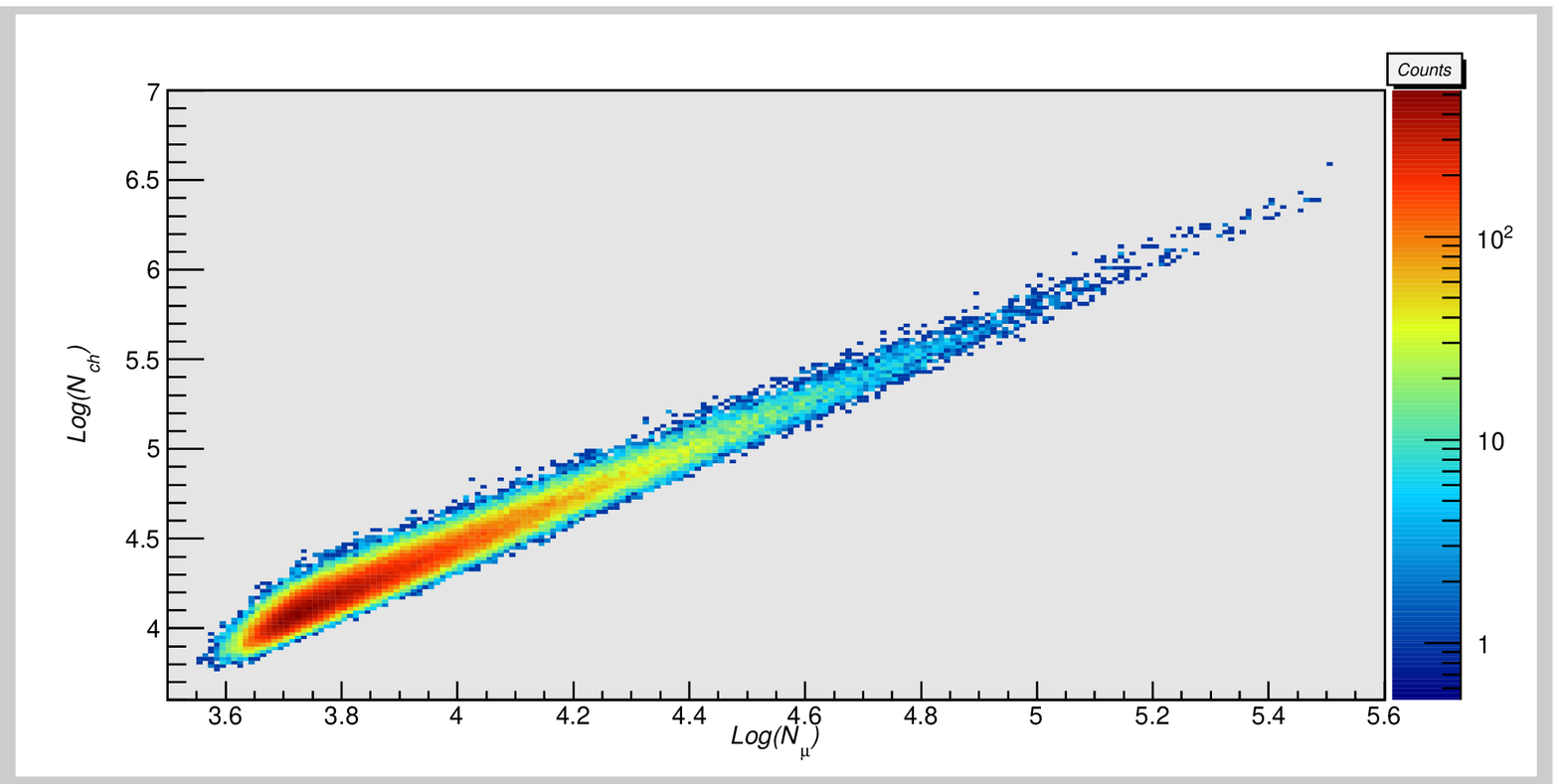}}
\caption{(Color online) 2-dimentional charged particles - muon spectrum for different composition scenario around the knee}
\end{figure*}

\section{Discussion and Conclusion}

The knee of the primary energy spectrum has long been inferred from the break in shower size spectrum of cosmic ray EAS at certain shower size corresponding to few PeV primary energy. Few authors, particularly Stenkin, however, objected the existence of the knee in the primary energy spectrum noting that the muon size spectrum of cosmic ray EAS does not show any prominent break against the expectations. 

It is found from the present analysis that the EAS-TOP observations on total charged particle and muon spectra consistently infer a knee in the primary energy spectrum provided the primary is pure unchanging iron whereas no consistent primary spectrum emerges from simultaneous use of the KASCADE observed total charged particle and muon spectra. 

It is further found from Monte Carlo simulation results that for pure unchanging proton or iron primaries the difference in spectral slopes below and above the knee of the size spectrum is larger for muon spectrum than the electron spectrum. However, when mass composition changes across the knee the situation becomes quite complex. In such a situation estimation of $\beta$ properly is problamatic, particularly for total charged particle spectrum. The $\beta_{ch}$ and the position of the knee depend on primary composition both below and above the knee of the primary energy spectrum when the data points close to the knee in the size spectra are incorporated to determine them. A different choice of data points may change the overall slope considerably. For instance in the simple situation where proton and Fe are the dominating component below and above the knee of the primary energy spectrum, the contribution of Fe, which gives a comparative lower total number of charged particles, leads to a flatter shower size spectrum below the knee, unless the points closed to the knee in the size spectrum are totally ignored to evaluate the slopes. On the other hand iron induced EAS contains comparatively larger number of muons. Hence the slopes of the muon size spectrum does not alter much for the stated changing composition scenario but there will be a mismatch in the flux at the knee of the muon size spectrum. Non observation of any break in flux level at the knee position of the muon size spectrum in any experiment suggests that there is no abrupt change in primary composition across the knee; the composition either changes slowly above the knee or it changes from a lighter dominating mixed composition to heavier dominated mixed composition without appreciable change in average primary mass.   

We thus conclude that though the derivation of the size spectrum from observed data looks to be rather straight forward process, but in practice it is a quite complex issue, particularly owing to the uncertainty in primary mass composition. The simultaneous use of the measured EAS total charged particle and muon size spectra to infer the primary energy spectrum is certainly a better approach but it requires a careful and experiment specific analysis. The two-dimensional differential spectrum contents substantially higher information than those of two one-dimensional ones and hence used to infer primary spectrum and composition but one dimensional spectra also carry important and exclusive signatures about primary energy spectrum and composition which should also be accommodated to get reliable information about cosmic ray primaries.


\begin{thebibliography}{99}

\bibitem{t1}	C. E. Fichtel, and J. Linsley, Astrophys. J. \textbf{300}, 474 (1986)
\bibitem{t2}	V. L. Ginzburg, and S. I. Syrovatskii,  1964, The Origin of Cosmic Rays, Macmillan, NewYork.
\bibitem{t3}	A. D. Erlykin, A. W. Wolfendale, J. Phys. G: Nucl. Part. Phys. \textbf{23}, 979 (1997).
\bibitem{t4}	B. Bijay and A. Bhadra, Res. Astron. Astrophys. (to appear) (2015); eprint arXiv:1412.0818.
\bibitem{t5}	G. V. Kulikov, G. B. Khristiansen, JETP, \textbf{35}, 441 (1959).
\bibitem{t6}	S. I. Nikolsky, and V. A. Romachin, Physics of Atomic Nuclei, \textbf{63}, 1799 (2000).
\bibitem{t7}	D. Kazanas and A. Nicolaidis,  eprint arXiv:astro-ph/0103147 (2001).
\bibitem{t8}	Yu. V. Stenkin, Mod. Phys. Lett. {\bf A18} 1225 (2003) .
\bibitem{t9}  Yu. V. Stenkin, Nucl. Phys. B (Proc. Suppl.) \textbf{151}, 65 (2006)
\bibitem{t10}	D M Gromushkin et al., J. Phys. (Conf. Series) \textbf{409}, 012044 (2013).
\bibitem{t11}	D. Heck, J. Knapp, J. N. Capdevielle, G. Schatz and T. Thouw, Forschungszentrum Karlsruhe Report No. FZKA 6019, (1998).
\bibitem{t12}	J. Matthews, Astropart. Phys. {\bf 22} 387 (2005).
\bibitem{t13}	J. R. Hoerandel, Mod.Phys.Lett.A {\bf 22} 1533 (2007)
\bibitem{t14}	N. N. Kalmykov, S. S. Ostapchenko and A. I. Pavlov, Nucl. Phys. B, Proc. Suppl. {\bf 52} 17 (1997).
\bibitem{t15}	M.~Bleicher et al., J. Phys. G {\bf 25} 1859 (1999).
\bibitem{t16}	K.~Werner, F.~M.~Liu and T.~Pierog, Phys. Rev. C {\bf 74} 044902 (2006).
\bibitem{t17}	H. Fesefeldt, RWTH Aachen Report No. PITHA-85/02, (1985).
\bibitem{t18}	H. J. Drescher, M. Bleicher, S. Soff and H. St¨ocker, Astropart. Phys. \textbf{21} 87 (2004).
\bibitem{t19}	A. Bhadra, S. K. Ghosh, P. S. Joarder, A. Mukherjee and S. Raha, Phys. Rev. D \textbf{79} 114027 (2009).
\bibitem{t20} T. Antoni, et.al. (KASCADE collab.), Nucl. Instru. Meth. \textbf{513} 490 (2003)
\bibitem{t21} M. Aglietta et al. (EAS-TOP collab.) IL Nuovo Cim. 9C, 262 (1986)
\bibitem{t22} R. Glasstetter et al., Proc. Int. Cosmic Ray Conf. \textbf{6}, 157 (1997)
\bibitem{t23} T. Antoni et. al (KASCADE collab.), Astropart. Phys. \textbf{16} 373 (2002)
\bibitem{t24} G. Navarra et al (EAS-TOP collab.), Nucl. Phys. B (Proc. Suppl.) 60, 105 (1998)
\bibitem{t25} kcdc.ikp.kit.edu (KIT, Karlsruhe Institute of Technology)




\end{thebibliography}
\end{document}